\documentclass[preprint,12pt]{elsarticle}
\journal{Computer Physics Communications}

\usepackage{amsmath}
\usepackage[T1]{fontenc}
\usepackage{framed}

\usepackage[utf8]{inputenc}
\usepackage{url}
\usepackage{graphicx}

\newenvironment{code}
  {\tt\noindent\begin{framed}\begin{flushleft}}
  {\end{flushleft}\end{framed}\normalfont}

\begin{document}

\begin{frontmatter}

\title{FIRE 6.5: Feynman Integral Reduction with New Simplification Library}

\author[SRCC,MCM]{Alexander V.~Smirnov\corref{cor1}}
\ead{asmirnov80@gmail.com}

\author[HI]{Mao Zeng}
\ead{mao.zeng@ed.ac.uk}

\cortext[cor1]{Corresponding author}

\address[SRCC]{Research Computing Center, Moscow State University, \\ 119992 Moscow, Russia}
\address[MCM]{Moscow Center for Fundamental and Applied Mathematics, 119992, Moscow, Russia}
\address[HI]{Higgs Centre for Theoretical Physics, University of Edinburgh,\\
James Clerk Maxwell Building, Peter Guthrie Tait Road, Edinburgh, EH9 3FD,\\
United Kingdom}

\begin{abstract}
  FIRE is a program which performs integration-by-parts (IBP) reduction of Feynman integrals.
  Originally, the C++ version of FIRE relies on the computer algebra system Fermat by Robert Lewis to simplify rational functions.
  We present an upgrade of FIRE which incorporates a new library FUEL initially described in a separate publication, which enables a flexible choice of third-party computer algebra systems as simplifiers, as well as efficient communications with some of the simplifiers as C++ libraries rather than through Unix pipes. We achieve significant speedups for IBP reduction of Feynman integrals involving many kinematic variables, when using an open source backend based on FLINT newly added in this work, or the Symbolica backend developed by Ben Ruijl as a potential successor of FORM.
\end{abstract}
\end{frontmatter}
\newpage

{\bf PROGRAM SUMMARY}

\vspace{1cm}

\begin{small}
\noindent
    {\em Program title:} FIRE, version 6.5 (FIRE 6.5)\\
    {\em CPC Library link to program files:} (to be added by Technical Editor) \\
    {\em Developer's repository link:} \url{https://gitlab.com/feynmanintegrals/fire.git} \\
    {\em Code Ocean capsule:} (to be added by Technical Editor)\\
    {\em Licensing provisions:} GPLv2\\
    {\em Programming language:} {\tt Wolfram Mathematica} 8.0 or higher, {\tt C++17}\\
    {\em Supplementary material:} see linked repository for installation instructions\\
    {\em Journal reference of previous version:} https://doi.org/10.1016/j.cpc.2019.106877\\
    {\em Does the new version supersede the previous version?} Yes.\\
    {\em Reasons for the new version:} The new version no longer relies on a single computer algebra system, {\tt Fermat} [1], but instead allows a flexible choice of several systems, some of which offer higher performance, especially when the number of variables is large.\\
    {\em Summary of revisions:} A new library {\tt FUEL} [2] is used as a core component of the new version of {\tt FIRE} to access different computer algebra systems as simplifiers of rational function expressions. Since the first release of {\tt FUEL} described elsewhere, {\tt FUEL} version 1.0 here has been enhanced with a new backend based on the open source library {\tt FLINT} [3] that provides highly performant simplification of rational functions.\\
    {\em Nature of problem:}
    Feynman integrals of a given family are reduced to a finite set of master integrals, by solving linear equations arising from integration by parts, using Gaussian elimination. The coefficient of the linear equations are generally rational functions in kinematic variables and the spacetime dimension, and the simplification of such rational functions during Gaussian elimination is a key task that is improved in this upgrade of {\tt FIRE}.\\
    {\em Solution method:}
    Computer algebra systems with state-of-the-art capabilities for polynomial GCD computations are used as simplification backends, or simplifiers in short. Due to the design of {\tt FIRE}, text strings are used as the exchange format for rational functions before and after simplification. A fast {\tt C++} parser is written to parse strings into the internal format of an external simplifer, {\tt FLINT} [3], with state-of-the-art performance for multivariate polynomial calculations. Similarly, the simplifier {\tt Symbolica} [4] has high performance in both GCD computations and parsing, and has been integrated into FIRE.\\
{\em References:} 
{\\}
[1] \url{https://home.bway.net/lewis/}, free--ware with some restrictions;
{\\}
[2] \url{https://doi.org/10.26089/NumMet.v24r425}, open source;
{\\}
[3] \url{https://flintlib.org/}, open source;
{\\}
[4] \url{https://symbolica.io/}, commerical software with free licenses for students and hobbyists

\end{small}

\newpage

\section{Introduction}
Evaluating Feynman integrals underlies perturbative calculations in quantum field theory. Integration-by-parts (IBP) reduction \cite{Chetyrkin:1981qh} is a standard step in evaluating Feynman integrals and expresses all integrals of a given family, with a common set of propagators, in terms of a finite number of master integrals. IBP reduction exploits the fact that total derivatives integrate to zero in dimensional regularization, which generates linear relations, referred to as \emph{integration-by-parts equations}, between different integrals of a given family. Typically, master integrals are chosen to be ``simple'' integrals where propagators are not raised to high powers and numerators are low-degree monomials.

In all but the simplest applications, one must consider non-obvious linear sums of IBP equations to accomplish the reduction. One systematic approach is the Laporta algorithm \cite{Laporta:2000dsw}. Some publicly available implementations of the Laporta algorithm include {\tt AIR} \cite{AIR}, {\tt Reduze} \cite{Reduze}, {\tt LiteRed} \cite{LiteRed}, {\tt FIRE} \cite{Smirnov:2013dia, Smirnov:2014hma, FIRE6}, and {\tt Kira} \cite{Maierhofer:2017gsa, Kira, Klappert:2020nbg}. FIRE has been developed by one of the current authors and collaborators over the past decade and has found many applications in cutting-edge calculations in high energy physics. The Laporta algorithm generates all IBP equations under a certain cutoff of the most complicated integral involved, before solving the system of linear equations using Gaussian elimination. The linear equations have coefficients that are polynomials in the spacetime dimension $d$ and kinematic variables (e.g.\ Mandelstam variables and mass parameters), and the coefficients become rational functions of the same set of variables during Gaussian elimination. These rational functions need to be simplified (usually into single fractions whose numerators and denominators have no nontrivial polynomial GCD) throughout the process of Gaussian elimination to prevent the expressions from exploding in size.

An alternative to the Laporta algorithm is deriving reduction rules with symbolic dependence on the powers of propagators and irreducible scalar products, through heuristic algorithms, e.g.\ implemented in {\tt LiteRed} \cite{Lee:2012cn}, or through Groebner bases \cite{Tarasov:2004ks, Gerdt:2005qf, Smirnov:2005ky, Smirnov:2006tz, Smirnov:2006wh, Lee:2008tj, Barakat:2022qlc}. The reduction rules express complicated integrals in terms of simpler integrals (according to some ordering), and the rules generally need to be applied repeatedly to express a given integral in terms of master integrals. Therefore the simplification of rational functions in intermediate steps is again an essential computational task. The approach of syzygy equations \cite{Gluza:2010ws, Schabinger:2011dz, Ita:2015tya, Larsen:2015ped, Abreu:2017xsl, Abreu:2017hqn, Bohm:2018bdy, Bendle:2019csk, Wu:2023upw} reduces the size of the linear system, while the approach of block triangular forms \cite{Liu:2018dmc, Guan:2019bcx} makes the linear system more structured, but the need to simplify rational functions in intermediate steps remains.

Simplification of rational functions is a functionality provided by \emph{computer algebra systems}. Some IBP programs are written in a computer algebra system already, so this functionality can be accessed naturally. For example, AIR is written in Maple, and FIRE has an alternative {\tt Wolfram Mathematica} variant besides the much faster {\tt C++} variant. However, many of the high-performance IBP programs are written in a general-purpose programming language. This requires interfacing an external computer algebra system as either a separate program or as a library. For example, {\tt Reduze}, {\tt Kira}, and (the preferred variant of) {\tt FIRE} are primarily written in C++. All these three programs (until the new version of FIRE presented here) use the Fermat computer algebra system by default for simplifying rational functions.\footnote{The authors of {\tt Reduze} explored using {\tt GiNaC} \cite{Bauer:2000cp} instead but found the speed to uncompetitive against Fermat at the time of writing of their paper \cite{Reduze}. Ref.~\cite{Bendle:2019csk} uses {\tt Singular} \cite{Singular} for simplification of rational functions.} Fermat is a command line program that takes text input and generates text output, and the aforementioned IBP programs communicate with Fermat via bi-directional Unix pipes, mimicking how a human user would interact with a command-line program. Specifically, previous versions of FIRE communicates with Fermat using the {\tt gateToFermat} library by M.\ Tentukov.

A recent paper \cite{Mokrov:2023vva} by the current authors has presented a C++ library, {\tt FUEL}, which provides an uniform interface to many computer algebra systems as simplification backends, or \emph{simplifiers} in short. This allows a flexible choice of simplifiers, some of which are found to out-perform Fermat for various IBP reduction problems. Additionally, {\tt FUEL} can communicate with some simplifiers as C++ libraries, which enhances performance by removing the overhead of pipes.

In this paper, we present a new version of {\tt FIRE} which uses {\tt FUEL} for simplification of rational functions. The user interface is largely unchanged from previous versions of FIRE, except that users are now provided with a choice of third-party computer algebra systems as the simplifier. The build system has been updated to automatically handle the installation of some of the third-party software. We present FIRE benchmarks to compare the three top-performing simplifier backends: {\tt Fermat} \cite{FERMAT}, {\tt FLINT} \cite{flint}, and {\tt Symbolica} \cite{Symbolica}. The benchmark IBP reduction problems involve a variety of loop orders and numbers of variables, and include runs based on either the Laporta algorithm or {\tt LiteRed} reduction rules. In particular, the ability to use {\tt FLINT} for our purpose is first presented in this paper. Other new features added to FIRE since the announcement of version 6.0 \cite{FIRE6}, covered in other publications, includes balanced reconstruction of rational functions \cite{Belitsky:2023qho} and improving the choice of master integrals \cite{Smirnov:2020quc}. The latter feature has been separated into a standalone package after initial inclusion as part of FIRE. (See also Ref.~\cite{Usovitsch:2020jrk} for related work.)

We mention in passing that an entirely different approach is reconstructing rational functions from finite-field numerical evaluations \cite{vonManteuffel:2014ixa, Peraro:2016wsq, Klappert:2019emp, Peraro:2019svx, Laurentis:2019bjh, DeLaurentis:2022otd, Magerya:2022hvj, Belitsky:2023qho}, bypassing the need to simplify rational functions in intermediate steps. A related technique is the further simplification of rational functions by multivariate partial fractioning \cite{Abreu:2019odu, Heller:2021qkz}, \emph{after} analytic results for IBP reduction tables or complete loop amplitudes have already been obtained.

\section{Performance considerations in rational function simplification}
The complete list of supported simplifiers included in {\tt FUEL}, and hence available in the new version of {\tt FIRE}, can be found on the {\tt README} on \\
\url{https://gitlab.com/feynmanintegrals/fuel.git}.\\
In this section, we focus on {\tt FLINT} and {\tt Symbolica}, which achieve competitive performance across a wide range of IBP reduction problems when used with {\tt FIRE}. In Ref.~\cite{Mokrov:2023vva} which presented {\tt FUEL}, the Nemo computer algebra system \cite{Nemo} was investigated as a simplifier, which provided much of the prototyping for our subsequent use of {\tt FLINT} in {\tt FUEL} 1.0 described here. Nemo is a package for the Julia programming language, and the underlying engine for polynomial computations is the C library {\tt FLINT}. By switching to using {\tt FLINT} directly, we find it easier to gain control over performance. We use recent development versions of {\tt FLINT} which contain an implementation of multivariate rational function arithmetic on top of its more established polynomial functionality.\footnote{Originally, multivariate rational functions are implemented in the {\tt Calcium} library \cite{Calcium}, before being merged into the development repository of {\tt FLINT} earlier this year. The relevant functionality is available in the header file {\tt fmpz\string_mpoly\string_q.h}.} {\tt Symbolica} \cite{Symbolica} is being developed by Ben Ruijl as a modernized alternative to FORM \cite{Ruijl:2017dtg}, and the use of Symbolica  in {\tt FUEL} was already investigated in Ref.~\cite{Mokrov:2023vva}. The author of {\tt Symbolica} has further optimized the software for our purposes since then.

In the subsections below, we briefly describe some of the performance considerations that have been taken into account in the {\tt FLINT} and {\tt Symbolica} backends.
\subsection{Overhead in string representations of expressions}
In {\tt FIRE} and {\tt FUEL}, rational functions are stored as strings in the common mathematical notation, e.g.~{\tt (2*x-y)/(x+y\string^2)}. Though not without drawbacks, this is a convenient choice, since (1) this format is accepted by almost any third-party computer algebra system; (2) database systems, used to store intermediate results, commonly accept string data; (3) strings are used to exchange expressions between different worker processes running in parallel.

When {\tt FIRE} needs to simplify an expression, it is sent as a string to the external simplifier, which parses the string into an internal representation, performs simplifications, and prints out the simplified expression. As pointed out in the initial paper for {\tt FUEL} \cite{Mokrov:2023vva}, an external simplifier must have excellent performance in not only the simplification itself but also parsing and printing to achieve competitive performance. For example, the aforementioned reference finds Maple to be among the fastest simplifiers in contrived tests that minimize the overhead of parsing and printing, but Maple performs poorly for realistic FIRE workloads due to the overhead.

{\tt FLINT} has a built-in parser for polynomials but not rational functions. We have written a fast {\tt C++} parser to convert strings representing rational functions into {\tt FLINT}'s internal rational function type. The parsing algorithm is a variant of the well-known shunting yard algorithm for parsing mathematical expressions while taking operator precedence into account. {\tt Symbolica} also has a fast parser for rational function expressions. The good parsing performance is essential for the good performance of {\tt FLINT} and {\tt Symbolica} backends in the simpler IBP reduction problems tested in Section \ref{sec:benchmarks}, since the parsing overhead is usually significant in this situation \cite{Mokrov:2023vva}.

\subsection{Re-evaluation cost}
Since {\tt FIRE} stores rational functions as strings, binary arithmetic operators are implemented by concatenating the operands with the appropriate operator characters. For example, to add two rational functions that were previously simplified, written schematically as {\tt (poly1)/(poly2)} and {\tt (poly3)/(poly4)}, {\tt FIRE} forms the string\\
{\tt (poly1)/(poly2) + (poly3)/(poly4)}\\
and sends it (via the {\tt FUEL} library) to the simplifier. However, now the simplifier does not make assumptions about the input and attempts to simplify the two rational functions again before adding them. For example, the simplifier will attempt to find the polynomial GCD of {\tt poly1} and {poly2} and divide both polynomials by the GCD. This step can be avoided if it is known that the rational function is already simplified and the numerator and denominator have no nontrivial GCD.

To solve this problem, the {\tt FLINT} and {\tt Symbolica} backends have been programmed to print out simplified rational functions in the custom notation\\
{\tt [poly1,poly2]}
instead of {\tt (poly1)/(poly2)}, to indicate that the rational function has already been simplified. When such strings are inserted into further computations, the parser will directly produce a rational function given by the numerator {\tt poly1} and denominator {\tt poly2}, without attempting a redundant polynomial GCD computation. When a simplified expression is a single polynomial rather than a rational function, the custom notation\\
{\tt [poly]}
is used.\footnote{When this notation is encountered in further computations, the {\tt FLINT} backend uses a polynomial parser that does not handle rational functions, and the Symbolica backend uses a dedicated parser for polynomials that are known to have been expanded.}

At the end of a FIRE run, all such non-conventional strings are converted back to the conventional mathematical format, for the results (IBP reduction tables) to be consumed by end users, e.g.\ using Wolfram Mathematica.

\section{Benchmarks}
\label{sec:benchmarks}
All the benchmarks are run on two machines whose characteristics are summarized in Table \ref{tab:machines}.
\begin{table}
  \footnotesize
    \centering
    \begin{tabular}{|c|c|c|c|c|c|}
      \hline
      Machine No. & CPU & No.\ of cores & Frequency & Cache Size & RAM \\ \hline
      1 & Intel Core i5-9500 & 6 & 3.00 GHz & 9216 KB & 16 GB \\ \hline
      2 & Intel Core i7-8550U & 4 & 1.80 GHz & 8192 KB & 16 GB \\ \hline
    \end{tabular}
    \caption{Characteristics of the two machines used to run benchmark tests}
    \label{tab:machines}
\end{table}
Each benchmark tests the time taken by a FIRE run to perform an IBP reduction task. The FIRE configurations allow up to 4 sector worker threads and up to 4 simplifier threads to run in parallel. The results for each benchmark tests are displayed as two numbers to indicate the time taken on Machine 1 and Machine 2, respectively.\footnote{The versions of the simplifiers are as follows. {\tt FLINT}: \url{https://github.com/wbhart/flint2.git}, commit hash {\tt 283e51}, Fri Sep 8 2023. {\tt Symbolica}: \url{https://github.com/benruijl/symbolica.git}, commit hash {\tt 10c562}, Oct 7 2023. Fermat: version 5.17. The later version Fermat 7.0 has certain improvements, but we do not see any noticeable speedup for the benchmarks in this paper with up to 5 variables, possibly due to overhead such as parsing.}
\subsection{Nonplanar vertex diagram}
\begin{figure}
  \centering
  \includegraphics[width=0.4\textwidth]{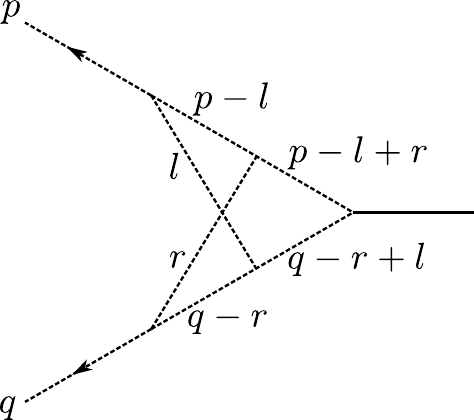}
  \caption{The nonplanar vertex diagram with two loops.}
  \label{fig:vertex}
\end{figure}

The nonplanar vertex diagram is shown in Fig.~\ref{fig:vertex}
We set
\begin{equation}
  p^2 = q^2 = 0, \quad p\cdot q = -1/2,
\end{equation}
without nontrivial kinematic variables, so only the spacetime dimension $d$ will be involved in solving the IBP system. As in an existing example in previous versions of FIRE, reduction rules produced by LiteRed \cite{Lee:2012cn} are used. The only irreducible scalar product is
\begin{equation}
  (l-r)^2 \, .
\end{equation}
We test a simple IBP problem of reducing the vertex integral with 1 power of the above ISP, as well as a harder IBP problem of reducing the vertex integral with the same ISP raised to the 20th power. The execution times are in Table~\ref{tab:vertextest}. To reduce noise, every run is repeated 10 times, and the median value of the execution time is taken. For reducing a rank-20 integral, {\tt FLINT} and {\tt Symbolica} show moderate speedups over {\tt Fermat}. Note that the {\tt FLINT} backend could be further optimized for univariate rational functions involved in this IBP reduction problem, since we always interface with multivariate rational functions provided by {\tt FLINT}, which is sub-optimal for the special univariate case.
\begin{table}
  \footnotesize
    \centering
    \begin{tabular}{|c|c|c|c|c|}
      \hline
      Simplifier& rank-1 time (s) & rank-20 time (s)  \\ \hline
      FLINT & 0.70, \ 0.93 & 11, \ 18 \\ \hline
      Symbolica & 0.65, \ 0.84 & 13, \ 18 \\ \hline
      Fermat & 0.65, \ 0.91 & 14, 24 \\ \hline
    \end{tabular}
    \caption{Time taken for FIRE to perform the IBP reduction for the nonplanar vertex diagram, with rank-1 or rank-20 numerators, for three different simplifier backends. Every result is shown as two numbers, indicating the time measured on the two different machines.}
    \label{tab:vertextest}
\end{table}

\subsection{Two-loop massless double box}
\begin{figure}
  \centering
  \includegraphics[width=0.4\textwidth]{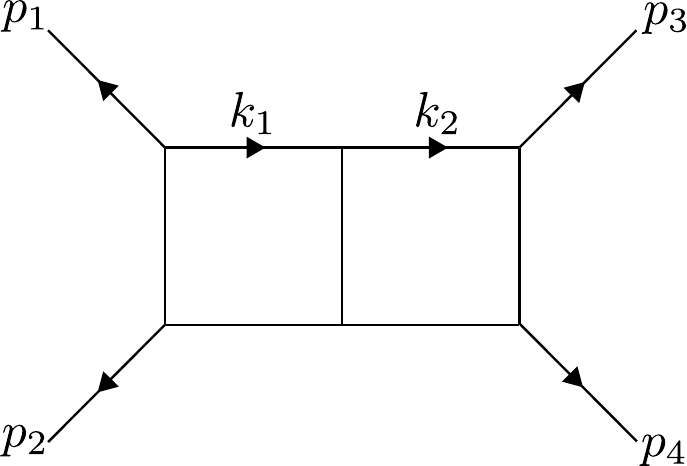}
  \caption{The double box diagram.}
  \label{fig:doublebox}
\end{figure}

The double box diagram is shown in Fig.~\ref{fig:doublebox}.
The kinematic variables are
\begin{equation}
  (p_1 + p_2)^2 = s, \quad (p_1+p_3)^2 = t \, .
\end{equation}
The two irreducible scalar products are chosen as 
\begin{equation}
  (k_2+p_1)^2, \quad (k_1-p_3)^2 \, .
\end{equation}
We test an easy problem of the reduction of the integral with a rank-2 numerator (i.e.~degree 2 in the irreducible scalar products), $(k_2+p_1)^2 (k_1-p_3)^2$, as well as a harder problem of the reduction of the integral with a rank-8 numerator, $[(k_2+p_1)^2]^4 [(k_1-p_3)^2]^4$. The easier problem is one of the examples included in existing versions of FIRE. Symmetry rules for master integrals as well as preferences for master integrals are supplied to FIRE. The execution times are shown in Table.~\ref{tab:dboxtest}. For the rank-2 integral, every run is repeated 10 times, and the median value of the execution time is taken, while the rank-8 integral is tested with single runs. For reducing a rank-8 integral, {\tt FLINT} and {\tt Symbolica} show significant speedups over Fermat.
\begin{table}
  \footnotesize
    \centering
    \begin{tabular}{|c|c|c|c|c|}
      \hline
      Simplifier& rank-2 time (s) & rank-8 time (1000 s) \\ \hline
      FLINT & 7.5, \ 10.8 & 0.13, \ 0.30 \\ \hline
      Symbolica & 6.7, \ 9.2 & 0.13, \ 0.26 \\ \hline
      Fermat & 7.9, \ 14.9 & 0.25, \ 0.48  \\ \hline
    \end{tabular}
    \caption{Time taken for FIRE to perform the IBP reduction for the massless double box diagram, with rank-2 or rank-8 numerators, for three different simplifier backends. Every result is shown as two numbers, indicating the time measured on the two different machines.}
    \label{tab:dboxtest}
\end{table}

\subsection{Four-loop diagram for classical two-body interaction}
\begin{figure}
  \centering
  \includegraphics[width=0.6\textwidth]{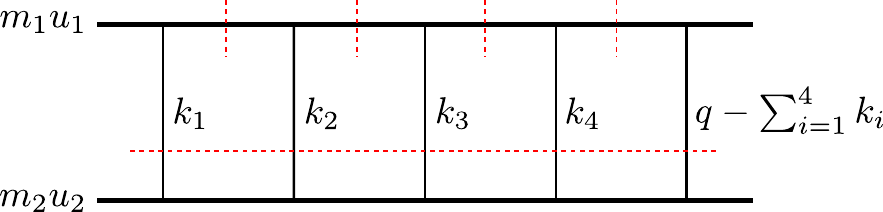}
  \caption{The four-loop iterated ladder diagram for classical two-body scattering. The red dashed lines indicate the nine propagators that are cut in the IBP reduction calculation.}
  \label{fig:soft_quad_box}
\end{figure}

This is an example from Ref.~\cite{Bern:2023ccb} on the classical scattering of charged bodies in electrodynamics at the 5th order in $\alpha$. One of the diagrams involved is shown in Fig.~\ref{fig:soft_quad_box}.
We work in the kinematic limit that the massless vertical lines in the middle are much softer than the momenta carried by the massive horizontal lines on the top and bottom. As a result, all the massive propagators are expanded to leading order as linearized propagators related to the normalized velocities $u_1$ and $u_2$ of the massive incoming momenta on the left. The propagators are
\begin{align}
  &  k_1^2, \, k_2^2, \, k_3^2, \, k_4^2, \, (q-k_1-k_2-k_3-k_4)^2, \, 2 u_1 \cdot k_1, \, 2 u_1 \cdot (k_1+k_2),\,  2 u_1 \cdot (k_1+k_2+k_3), \nonumber \\
  & 2 u_1 \cdot (k_1+k_2+k_3+k_4), \, -2 u_2 \cdot k_1, \, -2 u_2 \cdot (k_1+k_2), \, -2 u_2 \cdot (k_1+k_2+k_3), \nonumber \\
  & -2 u_2 \cdot (k_1+k_2+k_3+k_4) \, .
\end{align}
We have $u_1^2=u_2^2=1$ and set $q^2=-1$. The only nontrivial kinematic variable is $y= u_1 \cdot u_2$. We reduce a rank-2 tensor integral with numerator $(k_3 \cdot k_4)^2$ on a nine-particle cut indicated by red dashed lines in Fig.~\ref{fig:soft_quad_box}. An integral is set to zero if any of the nine propagators is canceled by a numerator. Full results for the IBP reduction can be subsequently obtained by merging IBP reduction results on a full set of spanning cuts (see e.g.~\cite{Larsen:2015ped}). The execution times are shown in Table~\ref{tab:4looptest}. Again, {\tt FLINT} and {\tt Symbolica} demonstrate speedups over Fermat. Much more dramatic speedups will be seen in the next example.
\begin{table}
  \footnotesize
    \centering
    \begin{tabular}{|c|c|}
      \hline
      Simplifier& Time (1000s) \\ \hline
      FLINT & 0.45, \ 0.97 \\ \hline
      Symbolica & 0.45, \ 0.94 \\ \hline
      Fermat & 0.82, \ 1.60 \\ \hline
    \end{tabular}
    \caption{Time taken for FIRE to perform the IBP reduction for the soft-expanded 4-loop iterated ladder diagram with a rank-2 numerator, for three different simplifier backends. Every result is shown as two numbers, indicating the time measured on the two different machines.}
    \label{tab:4looptest}
\end{table}

\subsection{Three-loop banana diagram}
\begin{figure}
  \centering
  \includegraphics[width=0.4\textwidth]{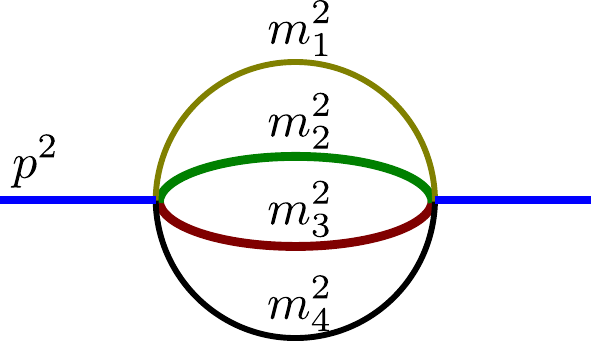}
  \caption{The three-loop banana diagram.}
  \label{fig:banana}
\end{figure}

The three-loop banana diagram is shown in Fig.~\ref{fig:banana}.
We consider the unequal mass case, with four different internal masses $m_1, m_2, m_3, m_4$. The external mass is set to a numerical value, $p^2=1$. We reduce a tensor integral whose numerator is a dot product between the loop momentum for the $m_2$ line and the loop momentum for the $m_3$ line. The execution times are shown in Table~\ref{tab:bananatest}. Unlike the previous benchmarks, this one has an unusually large number of kinematic scales. In this case, {\tt FLINT} and {\tt Symbolica} backends offer dramatic speedups, up to about 9 times, over the {\tt Fermat} backend.\footnote{\label{note1}Though {\tt FLINT} is the fastest for this benchmark, we have also encountered five-variable IBP reduction problems for which {\tt Symbolica} is the fastest \cite{Mokrov:2023vva}, with both {\tt FLINT} and {\tt Symbolica} leading Fermat considerably according to re-testing.}
\begin{table}[h]
  \footnotesize
    \centering
    \begin{tabular}{|c|c|}
      \hline
      Simplifier& Time (1000s) \\ \hline
      FLINT & 0.62, \ 1.01 \\ \hline
      Symbolica & 1.02, \ 1.66 \\ \hline
      Fermat & 5.36, \ 9.23 \\ \hline
    \end{tabular}
    \caption{Time taken for FIRE to perform the IBP reduction for the 3-loop banana diagram with unequal internal masses, with a rank-1 numerator, for three different simplifier backends. Every result is shown as two numbers, indicating the time measured on the two different machines.\protect\footnotemark[5]}
    \label{tab:bananatest}
\end{table}

\section{Installation and Usage}
Most of the installation and usage instructions are unchanged from the original paper describing {\tt FIRE} 6.0 \cite{FIRE6}, which should be the read by first-time users. Here we only describe the new features introduced in this {\tt FIRE} upgrade.

\subsection{Installation}
The installation instructions are similar to {\tt FIRE} 6.0 \cite{FIRE6}, except that the {\tt configure} script allows additional optional arguments for installing third-party simplifiers to be used by {\tt FIRE} through {\tt FUEL}. For example, to compile {\tt FLINT} and {\tt Symbolica} backends, in addition to the bundled Fermat backend, run
\begin{code}
./configure --enable\string_flint --enable\string_symbolica
\end{code}
As in {\tt FIRE} 6.0, the {\tt ZStandard} library can be used for compressed database storage, and the {\tt tcmalloc} library can be used for faster memory allocations. This changes the above command to
\begin{code}
./configure --enable\string_flint --enable\string_symbolica --enable\string_zstd --enable\string_tcmalloc
\end{code}
Then the dependencies can be compiled by
\begin{code}
  make dep
\end{code}
while {\tt FIRE} itself can be compiled by
\begin{code}
  make
\end{code}
Since both the {\tt FLINT} library and the finite-field part of {\tt FIRE} depend on {\tt GMP} and {\tt MPFR} libraries for arbitrary-precision arithmeetic, the source code for the latter two libraries have been bundled and will be built during the {\tt make dep} step. For convenience, we have optionally supplied prebuilt binaries of these two libraries which work for most current Linux distributions that use {\tt glibc} as the C library, such as Ubuntu 22.04 and CentOS 7. The use of the prebuilt binaries can be enabled by the following extra options supplied to the command line of {\tt configure}:
\begin{code}
-{}-enable\string_prebuilt\string_gmp -{}-enable\string_prebuilt\string_mpfr
\end{code}
\subsection{Usage}
The only significant change in usage compared with {\tt FIRE} 6.0 is the possibility to specify the simplifier used. Currently, {\tt Fermat} is still the default simplifier. To specify a different simplifer, e.g.\ {\tt FLINT}, one can use the
\begin{code}
-{}-calc flint
\end{code}
command line option for {\tt FIRE} or the finite-field variant {\tt FIREp}. Alternatively, one can add the line
\begin{code}
\#calc    flint
\end{code}
to the configuration file. Similarly, one could replace {\tt flint} by the name of another simplifier in lower-case letters, e.g.~{\tt symbolica}, if desired. Please note that the user is responsible for acquiring the licenses of any proprietary simplifiers, such as {\tt Fermat} and {\tt Symbolica}.

To run the benchmarks in Section \ref{sec:benchmarks} with e.g.\ {\tt FLINT}, one can run the following commands from the {\tt FIRE6} directory:
\begin{code}
  bin/FIRE6 -c examples/v2l -{}-calc flint\\
  bin/FIRE6 -c examples/v2l\string_high -{}-calc flint\\
  bin/FIRE6 -c examples/doubleboxrp -{}-calc flint\\
  bin/FIRE6 -c examples/doubleboxrp\string_high -{}-calc flint\\
  bin/FIRE6 -c examples/softQuadrupleBox -{}-calc flint\\
  bin/FIRE6 -c examples/bananaUnequal -{}-calc flint
\end{code}
The necessary {\tt .start} and {\tt .config} files for running the benchmarks have been included in {\tt FIRE}'s git repository. The {\tt benchmark.sh} script in the {\tt FIRE6} directory runs the complete benchmarks in Section \ref{sec:benchmarks} for all the three simplifiers discussed.

\section{Conclusion}
We have presented the new version 6.5 of {\tt FIRE}, a program for integration-by-parts reduction of Feynman integrals. The main change is the use of the {\tt FUEL} library which provides a uniform interface for simplifying rational functions, with a flexible choice of third-party simplifiers as backends. We have presented benchmarks for a variety of IBP reduction problems, including both runs based on the Laporta algorithm and runs based on {\tt LiteRed} reduction rules. The open source {\tt FLINT} backend, newly added in this work, and the {\tt Symbolica} backend both achieve performance that is comparable with, or moderately better than, the default {\tt Fermat} backend for simple IBP reduction problems with a small number of kinematic scales and low-complexity integrals. {\tt FLINT} and {\tt Symbolica} start to significantly out-perform Fermat when the complexity of the integrals increases, as can be seen in the benchmark for the double box diagram with a rank-8 numerator and the four-loop diagram involved in classical two-body scattering in electrodynamics. The performance advantages of {\tt FLINT} and {\tt Symbolica} over {\tt Fermat} become especially prominent for IBP problems with a large number of kinematic scales, with a speedup of up to a factor 9 seen in the benchmark for the 3-loop banana diagram with four unequal internal masses.

In the future, we would like to explore further ideas for optimizing the simplification of rational functions in {\tt FIRE}, e.g.~by exploring the freedom to reorder summation terms, as already implemented in {\tt Kira} \cite{Maierhofer:2017gsa}. More complicated IBP reduction tasks involving many kinematic scales will be tested, given the promising results already seen. We would also like to apply {\tt FUEL} to the reconstruction of multivariate rational functions through the balanced reconstruction algorithm \cite{Belitsky:2023qho}.

\section{Acknowledgments}
The authors would like to thank Enrico Herrmann, Ben Ruijl, and Johann Usovitsch for insightful discussions and/or comments on the manuscript.
The work of AS was supported by the Russian Science Foundation under Agreement No.\
21-71-30003 (usage of the FUEL library in the public version of FIRE) and by the Ministry of Education and Science of the Russian
Federation as part of the program of the Moscow Center for Fundamental and Applied
Mathematics under Agreement No.\ 075-15-2022-284 (developing the upgrade of the FUEL library that works directly with simplifier libraries).
M.Z.’s work is supported in part by the U.K.\ Royal Society through Grant URF\textbackslash R1\textbackslash 20109. For the purpose of open access, the authors have applied a Creative Commons Attribution (CC BY) license to any Author Accepted Manuscript version arising from this submission.
\bibliographystyle{elsarticle-num-names}
\bibliography{fuelfire}

\begin{thebibliography}{52}
\expandafter\ifx\csname natexlab\endcsname\relax\def\natexlab#1{#1}\fi
\providecommand{\url}[1]{\texttt{#1}}
\providecommand{\href}[2]{#2}
\providecommand{\path}[1]{#1}
\providecommand{\DOIprefix}{doi:}
\providecommand{\ArXivprefix}{arXiv:}
\providecommand{\URLprefix}{URL: }
\providecommand{\Pubmedprefix}{pmid:}
\providecommand{\doi}[1]{\href{http://dx.doi.org/#1}{\path{#1}}}
\providecommand{\Pubmed}[1]{\href{pmid:#1}{\path{#1}}}
\providecommand{\bibinfo}[2]{#2}
\ifx\xfnm\relax \def\xfnm[#1]{\unskip,\space#1}\fi
\bibitem[{Chetyrkin and Tkachov(1981)}]{Chetyrkin:1981qh}
\bibinfo{author}{K.~G. Chetyrkin}, \bibinfo{author}{F.~V. Tkachov},
\newblock \bibinfo{title}{{Integration by Parts: The Algorithm to Calculate
  beta Functions in 4 Loops}},
\newblock \bibinfo{journal}{Nucl. Phys. B} \bibinfo{volume}{192}
  (\bibinfo{year}{1981}) \bibinfo{pages}{159--204}.
  \DOIprefix\doi{10.1016/0550-3213(81)90199-1}.
\bibitem[{Laporta(2000)}]{Laporta:2000dsw}
\bibinfo{author}{S.~Laporta},
\newblock \bibinfo{title}{{High precision calculation of multiloop Feynman
  integrals by difference equations}},
\newblock \bibinfo{journal}{Int. J. Mod. Phys. A} \bibinfo{volume}{15}
  (\bibinfo{year}{2000}) \bibinfo{pages}{5087--5159}.
  \DOIprefix\doi{10.1142/S0217751X00002159}.
  \href{http://arxiv.org/abs/hep-ph/0102033}{{\tt arXiv:hep-ph/0102033}}.
\bibitem[{Anastasiou and Lazopoulos(2004)}]{AIR}
\bibinfo{author}{C.~Anastasiou}, \bibinfo{author}{A.~Lazopoulos},
\newblock \bibinfo{title}{{Automatic integral reduction for higher order
  perturbative calculations}},
\newblock \bibinfo{journal}{JHEP} \bibinfo{volume}{07} (\bibinfo{year}{2004})
  \bibinfo{pages}{046}. \DOIprefix\doi{10.1088/1126-6708/2004/07/046}.
  \href{http://arxiv.org/abs/hep-ph/0404258}{{\tt arXiv:hep-ph/0404258}}.
\bibitem[{von Manteuffel and Studerus(2012)}]{Reduze}
\bibinfo{author}{A.~von Manteuffel}, \bibinfo{author}{C.~Studerus},
\newblock \bibinfo{title}{{Reduze 2 - Distributed Feynman Integral Reduction}}
  (\bibinfo{year}{2012}). \href{http://arxiv.org/abs/1201.4330}{{\tt
  arXiv:1201.4330}}.
\bibitem[{Lee(2014)}]{LiteRed}
\bibinfo{author}{R.~N. Lee},
\newblock \bibinfo{title}{{LiteRed 1.4: a powerful tool for reduction of
  multiloop integrals}},
\newblock \bibinfo{journal}{J. Phys. Conf. Ser.} \bibinfo{volume}{523}
  (\bibinfo{year}{2014}) \bibinfo{pages}{012059}.
  \DOIprefix\doi{10.1088/1742-6596/523/1/012059}.
  \href{http://arxiv.org/abs/1310.1145}{{\tt arXiv:1310.1145}}.
\bibitem[{Smirnov and Smirnov(2013)}]{Smirnov:2013dia}
\bibinfo{author}{A.~V. Smirnov}, \bibinfo{author}{V.~A. Smirnov},
\newblock \bibinfo{title}{{FIRE4, LiteRed and accompanying tools to solve
  integration by parts relations}},
\newblock \bibinfo{journal}{Comput. Phys. Commun.} \bibinfo{volume}{184}
  (\bibinfo{year}{2013}) \bibinfo{pages}{2820--2827}.
  \DOIprefix\doi{10.1016/j.cpc.2013.06.016}.
  \href{http://arxiv.org/abs/1302.5885}{{\tt arXiv:1302.5885}}.
\bibitem[{Smirnov(2015)}]{Smirnov:2014hma}
\bibinfo{author}{A.~V. Smirnov},
\newblock \bibinfo{title}{{FIRE5: a C++ implementation of Feynman Integral
  REduction}},
\newblock \bibinfo{journal}{Comput. Phys. Commun.} \bibinfo{volume}{189}
  (\bibinfo{year}{2015}) \bibinfo{pages}{182--191}.
  \DOIprefix\doi{10.1016/j.cpc.2014.11.024}.
  \href{http://arxiv.org/abs/1408.2372}{{\tt arXiv:1408.2372}}.
\bibitem[{Smirnov and Chuharev(2020)}]{FIRE6}
\bibinfo{author}{A.~V. Smirnov}, \bibinfo{author}{F.~S. Chuharev},
\newblock \bibinfo{title}{{FIRE6: Feynman Integral REduction with Modular
  Arithmetic}},
\newblock \bibinfo{journal}{Comput. Phys. Commun.} \bibinfo{volume}{247}
  (\bibinfo{year}{2020}) \bibinfo{pages}{106877}.
  \DOIprefix\doi{10.1016/j.cpc.2019.106877}.
  \href{http://arxiv.org/abs/1901.07808}{{\tt arXiv:1901.07808}}.
\bibitem[{Maierh\"ofer et~al.(2018)Maierh\"ofer, Usovitsch, and
  Uwer}]{Maierhofer:2017gsa}
\bibinfo{author}{P.~Maierh\"ofer}, \bibinfo{author}{J.~Usovitsch},
  \bibinfo{author}{P.~Uwer},
\newblock \bibinfo{title}{{Kira\textemdash{}A Feynman integral reduction
  program}},
\newblock \bibinfo{journal}{Comput. Phys. Commun.} \bibinfo{volume}{230}
  (\bibinfo{year}{2018}) \bibinfo{pages}{99--112}.
  \DOIprefix\doi{10.1016/j.cpc.2018.04.012}.
  \href{http://arxiv.org/abs/1705.05610}{{\tt arXiv:1705.05610}}.
\bibitem[{Maierh\"ofer and Usovitsch(2018)}]{Kira}
\bibinfo{author}{P.~Maierh\"ofer}, \bibinfo{author}{J.~Usovitsch},
\newblock \bibinfo{title}{{Kira 1.2 Release Notes}}  (\bibinfo{year}{2018}).
  \href{http://arxiv.org/abs/1812.01491}{{\tt arXiv:1812.01491}}.
\bibitem[{Klappert et~al.(2021)Klappert, Lange, Maierh\"ofer, and
  Usovitsch}]{Klappert:2020nbg}
\bibinfo{author}{J.~Klappert}, \bibinfo{author}{F.~Lange},
  \bibinfo{author}{P.~Maierh\"ofer}, \bibinfo{author}{J.~Usovitsch},
\newblock \bibinfo{title}{{Integral reduction with Kira 2.0 and finite field
  methods}},
\newblock \bibinfo{journal}{Comput. Phys. Commun.} \bibinfo{volume}{266}
  (\bibinfo{year}{2021}) \bibinfo{pages}{108024}.
  \DOIprefix\doi{10.1016/j.cpc.2021.108024}.
  \href{http://arxiv.org/abs/2008.06494}{{\tt arXiv:2008.06494}}.
\bibitem[{Lee(2012)}]{Lee:2012cn}
\bibinfo{author}{R.~N. Lee},
\newblock \bibinfo{title}{{Presenting LiteRed: a tool for the Loop InTEgrals
  REDuction}}  (\bibinfo{year}{2012}).
  \href{http://arxiv.org/abs/1212.2685}{{\tt arXiv:1212.2685}}.
\bibitem[{Tarasov(2004)}]{Tarasov:2004ks}
\bibinfo{author}{O.~V. Tarasov},
\newblock \bibinfo{title}{{Computation of Grobner bases for two loop propagator
  type integrals}},
\newblock \bibinfo{journal}{Nucl. Instrum. Meth. A} \bibinfo{volume}{534}
  (\bibinfo{year}{2004}) \bibinfo{pages}{293--298}.
  \DOIprefix\doi{10.1016/j.nima.2004.07.104}.
  \href{http://arxiv.org/abs/hep-ph/0403253}{{\tt arXiv:hep-ph/0403253}}.
\bibitem[{Gerdt and Robertz(2006)}]{Gerdt:2005qf}
\bibinfo{author}{V.~P. Gerdt}, \bibinfo{author}{D.~Robertz},
\newblock \bibinfo{title}{{A Maple package for computing Grobner bases for
  linear recurrence relations}},
\newblock \bibinfo{journal}{Nucl. Instrum. Meth. A} \bibinfo{volume}{559}
  (\bibinfo{year}{2006}) \bibinfo{pages}{215--219}.
  \DOIprefix\doi{10.1016/j.nima.2005.11.171}.
  \href{http://arxiv.org/abs/cs/0509070}{{\tt arXiv:cs/0509070}}.
\bibitem[{Smirnov and Smirnov(2006)}]{Smirnov:2005ky}
\bibinfo{author}{A.~V. Smirnov}, \bibinfo{author}{V.~A. Smirnov},
\newblock \bibinfo{title}{{Applying Grobner bases to solve reduction problems
  for Feynman integrals}},
\newblock \bibinfo{journal}{JHEP} \bibinfo{volume}{01} (\bibinfo{year}{2006})
  \bibinfo{pages}{001}. \DOIprefix\doi{10.1088/1126-6708/2006/01/001}.
  \href{http://arxiv.org/abs/hep-lat/0509187}{{\tt arXiv:hep-lat/0509187}}.
\bibitem[{Smirnov(2006)}]{Smirnov:2006tz}
\bibinfo{author}{A.~V. Smirnov},
\newblock \bibinfo{title}{{An Algorithm to construct Grobner bases for solving
  integration by parts relations}},
\newblock \bibinfo{journal}{JHEP} \bibinfo{volume}{04} (\bibinfo{year}{2006})
  \bibinfo{pages}{026}. \DOIprefix\doi{10.1088/1126-6708/2006/04/026}.
  \href{http://arxiv.org/abs/hep-ph/0602078}{{\tt arXiv:hep-ph/0602078}}.
\bibitem[{Smirnov and Smirnov(2006)}]{Smirnov:2006wh}
\bibinfo{author}{A.~V. Smirnov}, \bibinfo{author}{V.~A. Smirnov},
\newblock \bibinfo{title}{{S-bases as a tool to solve reduction problems for
  Feynman integrals}},
\newblock \bibinfo{journal}{Nucl. Phys. B Proc. Suppl.} \bibinfo{volume}{160}
  (\bibinfo{year}{2006}) \bibinfo{pages}{80--84}.
  \DOIprefix\doi{10.1016/j.nuclphysbps.2006.09.032}.
  \href{http://arxiv.org/abs/hep-ph/0606247}{{\tt arXiv:hep-ph/0606247}}.
\bibitem[{Lee(2008)}]{Lee:2008tj}
\bibinfo{author}{R.~N. Lee},
\newblock \bibinfo{title}{{Group structure of the integration-by-part
  identities and its application to the reduction of multiloop integrals}},
\newblock \bibinfo{journal}{JHEP} \bibinfo{volume}{07} (\bibinfo{year}{2008})
  \bibinfo{pages}{031}. \DOIprefix\doi{10.1088/1126-6708/2008/07/031}.
  \href{http://arxiv.org/abs/0804.3008}{{\tt arXiv:0804.3008}}.
\bibitem[{Barakat et~al.(2023)Barakat, Br\"user, Fieker, Huber, and
  Piclum}]{Barakat:2022qlc}
\bibinfo{author}{M.~Barakat}, \bibinfo{author}{R.~Br\"user},
  \bibinfo{author}{C.~Fieker}, \bibinfo{author}{T.~Huber},
  \bibinfo{author}{J.~Piclum},
\newblock \bibinfo{title}{{Feynman integral reduction using Gr\"obner bases}},
\newblock \bibinfo{journal}{JHEP} \bibinfo{volume}{05} (\bibinfo{year}{2023})
  \bibinfo{pages}{168}. \DOIprefix\doi{10.1007/JHEP05(2023)168}.
  \href{http://arxiv.org/abs/2210.05347}{{\tt arXiv:2210.05347}}.
\bibitem[{Gluza et~al.(2011)Gluza, Kajda, and Kosower}]{Gluza:2010ws}
\bibinfo{author}{J.~Gluza}, \bibinfo{author}{K.~Kajda}, \bibinfo{author}{D.~A.
  Kosower},
\newblock \bibinfo{title}{{Towards a Basis for Planar Two-Loop Integrals}},
\newblock \bibinfo{journal}{Phys. Rev. D} \bibinfo{volume}{83}
  (\bibinfo{year}{2011}) \bibinfo{pages}{045012}.
  \DOIprefix\doi{10.1103/PhysRevD.83.045012}.
  \href{http://arxiv.org/abs/1009.0472}{{\tt arXiv:1009.0472}}.
\bibitem[{Schabinger(2012)}]{Schabinger:2011dz}
\bibinfo{author}{R.~M. Schabinger},
\newblock \bibinfo{title}{{A New Algorithm For The Generation Of
  Unitarity-Compatible Integration By Parts Relations}},
\newblock \bibinfo{journal}{JHEP} \bibinfo{volume}{01} (\bibinfo{year}{2012})
  \bibinfo{pages}{077}. \DOIprefix\doi{10.1007/JHEP01(2012)077}.
  \href{http://arxiv.org/abs/1111.4220}{{\tt arXiv:1111.4220}}.
\bibitem[{Ita(2016)}]{Ita:2015tya}
\bibinfo{author}{H.~Ita},
\newblock \bibinfo{title}{{Two-loop Integrand Decomposition into Master
  Integrals and Surface Terms}},
\newblock \bibinfo{journal}{Phys. Rev. D} \bibinfo{volume}{94}
  (\bibinfo{year}{2016}) \bibinfo{pages}{116015}.
  \DOIprefix\doi{10.1103/PhysRevD.94.116015}.
  \href{http://arxiv.org/abs/1510.05626}{{\tt arXiv:1510.05626}}.
\bibitem[{Larsen and Zhang(2016)}]{Larsen:2015ped}
\bibinfo{author}{K.~J. Larsen}, \bibinfo{author}{Y.~Zhang},
\newblock \bibinfo{title}{{Integration-by-parts reductions from unitarity cuts
  and algebraic geometry}},
\newblock \bibinfo{journal}{Phys. Rev. D} \bibinfo{volume}{93}
  (\bibinfo{year}{2016}) \bibinfo{pages}{041701}.
  \DOIprefix\doi{10.1103/PhysRevD.93.041701}.
  \href{http://arxiv.org/abs/1511.01071}{{\tt arXiv:1511.01071}}.
\bibitem[{Abreu et~al.(2017)Abreu, Febres~Cordero, Ita, Jaquier, Page, and
  Zeng}]{Abreu:2017xsl}
\bibinfo{author}{S.~Abreu}, \bibinfo{author}{F.~Febres~Cordero},
  \bibinfo{author}{H.~Ita}, \bibinfo{author}{M.~Jaquier},
  \bibinfo{author}{B.~Page}, \bibinfo{author}{M.~Zeng},
\newblock \bibinfo{title}{{Two-Loop Four-Gluon Amplitudes from Numerical
  Unitarity}},
\newblock \bibinfo{journal}{Phys. Rev. Lett.} \bibinfo{volume}{119}
  (\bibinfo{year}{2017}) \bibinfo{pages}{142001}.
  \DOIprefix\doi{10.1103/PhysRevLett.119.142001}.
  \href{http://arxiv.org/abs/1703.05273}{{\tt arXiv:1703.05273}}.
\bibitem[{Abreu et~al.(2018)Abreu, Febres~Cordero, Ita, Page, and
  Zeng}]{Abreu:2017hqn}
\bibinfo{author}{S.~Abreu}, \bibinfo{author}{F.~Febres~Cordero},
  \bibinfo{author}{H.~Ita}, \bibinfo{author}{B.~Page},
  \bibinfo{author}{M.~Zeng},
\newblock \bibinfo{title}{{Planar Two-Loop Five-Gluon Amplitudes from Numerical
  Unitarity}},
\newblock \bibinfo{journal}{Phys. Rev. D} \bibinfo{volume}{97}
  (\bibinfo{year}{2018}) \bibinfo{pages}{116014}.
  \DOIprefix\doi{10.1103/PhysRevD.97.116014}.
  \href{http://arxiv.org/abs/1712.03946}{{\tt arXiv:1712.03946}}.
\bibitem[{B\"ohm et~al.(2018)B\"ohm, Georgoudis, Larsen, Sch\"onemann, and
  Zhang}]{Bohm:2018bdy}
\bibinfo{author}{J.~B\"ohm}, \bibinfo{author}{A.~Georgoudis},
  \bibinfo{author}{K.~J. Larsen}, \bibinfo{author}{H.~Sch\"onemann},
  \bibinfo{author}{Y.~Zhang},
\newblock \bibinfo{title}{{Complete integration-by-parts reductions of the
  non-planar hexagon-box via module intersections}},
\newblock \bibinfo{journal}{JHEP} \bibinfo{volume}{09} (\bibinfo{year}{2018})
  \bibinfo{pages}{024}. \DOIprefix\doi{10.1007/JHEP09(2018)024}.
  \href{http://arxiv.org/abs/1805.01873}{{\tt arXiv:1805.01873}}.
\bibitem[{Bendle et~al.(2020)Bendle, B\"ohm, Decker, Georgoudis, Pfreundt,
  Rahn, Wasser, and Zhang}]{Bendle:2019csk}
\bibinfo{author}{D.~Bendle}, \bibinfo{author}{J.~B\"ohm},
  \bibinfo{author}{W.~Decker}, \bibinfo{author}{A.~Georgoudis},
  \bibinfo{author}{F.-J. Pfreundt}, \bibinfo{author}{M.~Rahn},
  \bibinfo{author}{P.~Wasser}, \bibinfo{author}{Y.~Zhang},
\newblock \bibinfo{title}{{Integration-by-parts reductions of Feynman integrals
  using Singular and GPI-Space}},
\newblock \bibinfo{journal}{JHEP} \bibinfo{volume}{02} (\bibinfo{year}{2020})
  \bibinfo{pages}{079}. \DOIprefix\doi{10.1007/JHEP02(2020)079}.
  \href{http://arxiv.org/abs/1908.04301}{{\tt arXiv:1908.04301}}.
\bibitem[{Wu et~al.(2023)Wu, Boehm, Ma, Xu, and Zhang}]{Wu:2023upw}
\bibinfo{author}{Z.~Wu}, \bibinfo{author}{J.~Boehm}, \bibinfo{author}{R.~Ma},
  \bibinfo{author}{H.~Xu}, \bibinfo{author}{Y.~Zhang},
\newblock \bibinfo{title}{{NeatIBP 1.0, A package generating small-size
  integration-by-parts relations for Feynman integrals}}
  (\bibinfo{year}{2023}). \href{http://arxiv.org/abs/2305.08783}{{\tt
  arXiv:2305.08783}}.
\bibitem[{Liu and Ma(2019)}]{Liu:2018dmc}
\bibinfo{author}{X.~Liu}, \bibinfo{author}{Y.-Q. Ma},
\newblock \bibinfo{title}{{Determining arbitrary Feynman integrals by vacuum
  integrals}},
\newblock \bibinfo{journal}{Phys. Rev. D} \bibinfo{volume}{99}
  (\bibinfo{year}{2019}) \bibinfo{pages}{071501}.
  \DOIprefix\doi{10.1103/PhysRevD.99.071501}.
  \href{http://arxiv.org/abs/1801.10523}{{\tt arXiv:1801.10523}}.
\bibitem[{Guan et~al.(2020)Guan, Liu, and Ma}]{Guan:2019bcx}
\bibinfo{author}{X.~Guan}, \bibinfo{author}{X.~Liu}, \bibinfo{author}{Y.-Q.
  Ma},
\newblock \bibinfo{title}{{Complete reduction of integrals in two-loop
  five-light-parton scattering amplitudes}},
\newblock \bibinfo{journal}{Chin. Phys. C} \bibinfo{volume}{44}
  (\bibinfo{year}{2020}) \bibinfo{pages}{093106}.
  \DOIprefix\doi{10.1088/1674-1137/44/9/093106}.
  \href{http://arxiv.org/abs/1912.09294}{{\tt arXiv:1912.09294}}.
\bibitem[{Bauer et~al.(2002)Bauer, Frink, and Kreckel}]{Bauer:2000cp}
\bibinfo{author}{C.~W. Bauer}, \bibinfo{author}{A.~Frink},
  \bibinfo{author}{R.~Kreckel},
\newblock \bibinfo{title}{{Introduction to the GiNaC framework for symbolic
  computation within the C++ programming language}},
\newblock \bibinfo{journal}{J. Symb. Comput.} \bibinfo{volume}{33}
  (\bibinfo{year}{2002}) \bibinfo{pages}{1--12}.
  \DOIprefix\doi{10.1006/jsco.2001.0494}.
  \href{http://arxiv.org/abs/cs/0004015}{{\tt arXiv:cs/0004015}}.
\bibitem[{Decker et~al.(2018)Decker, Greuel, Pfister, and
  Sch\"onemann}]{Singular}
\bibinfo{author}{W.~Decker}, \bibinfo{author}{G.-M. Greuel},
  \bibinfo{author}{G.~Pfister}, \bibinfo{author}{H.~Sch\"onemann},
  \bibinfo{title}{{\sc Singular} {4-1-1} --- {A} computer algebra system for
  polynomial computations},
  \bibinfo{howpublished}{\url{http://www.singular.uni-kl.de}},
  \bibinfo{year}{2018}.
\bibitem[{Mokrov et~al.(2023)Mokrov, Smirnov, and Zeng}]{Mokrov:2023vva}
\bibinfo{author}{K.~Mokrov}, \bibinfo{author}{A.~Smirnov},
  \bibinfo{author}{M.~Zeng},
\newblock \bibinfo{title}{{Rational Function Simplification for
  Integration-by-Parts Reduction and Beyond}}  (\bibinfo{year}{2023}).
  \href{http://arxiv.org/abs/2304.13418}{{\tt arXiv:2304.13418}}.
\bibitem[{Lewis(????)}]{FERMAT}
\bibinfo{author}{R.~H. Lewis}, \bibinfo{title}{{Fermat: A Computer Algebra
  System for Polynomial and Matrix Computation}},
  \bibinfo{howpublished}{\url{http://home.bway.net/lewis/}}, ????
  \bibinfo{note}{Accessed: 2023-02-18}.
\bibitem[{team(2023)}]{flint}
\bibinfo{author}{T.~F. team}, \bibinfo{title}{{FLINT}: {F}ast {L}ibrary for
  {N}umber {T}heory}, \bibinfo{year}{2023}. \bibinfo{note}{Version 2.9.0,
  \url{https://flintlib.org}}.
\bibitem[{Ruijl(2023)}]{Symbolica}
\bibinfo{author}{B.~Ruijl}, \bibinfo{title}{{Symbolica}},
  \bibinfo{howpublished}{\url{https://symbolica.io/}}, \bibinfo{year}{2023}.
  \bibinfo{note}{Accessed: 2023-06-23}.
\bibitem[{Belitsky et~al.(2023)Belitsky, Smirnov, and
  Yakovlev}]{Belitsky:2023qho}
\bibinfo{author}{A.~V. Belitsky}, \bibinfo{author}{A.~V. Smirnov},
  \bibinfo{author}{R.~V. Yakovlev},
\newblock \bibinfo{title}{{Balancing act: Multivariate rational reconstruction
  for IBP}},
\newblock \bibinfo{journal}{Nucl. Phys. B} \bibinfo{volume}{993}
  (\bibinfo{year}{2023}) \bibinfo{pages}{116253}.
  \DOIprefix\doi{10.1016/j.nuclphysb.2023.116253}.
  \href{http://arxiv.org/abs/2303.02511}{{\tt arXiv:2303.02511}}.
\bibitem[{Smirnov and Smirnov(2020)}]{Smirnov:2020quc}
\bibinfo{author}{A.~V. Smirnov}, \bibinfo{author}{V.~A. Smirnov},
\newblock \bibinfo{title}{{How to choose master integrals}},
\newblock \bibinfo{journal}{Nucl. Phys. B} \bibinfo{volume}{960}
  (\bibinfo{year}{2020}) \bibinfo{pages}{115213}.
  \DOIprefix\doi{10.1016/j.nuclphysb.2020.115213}.
  \href{http://arxiv.org/abs/2002.08042}{{\tt arXiv:2002.08042}}.
\bibitem[{Usovitsch(2020)}]{Usovitsch:2020jrk}
\bibinfo{author}{J.~Usovitsch},
\newblock \bibinfo{title}{{Factorization of denominators in
  integration-by-parts reductions}}  (\bibinfo{year}{2020}).
  \href{http://arxiv.org/abs/2002.08173}{{\tt arXiv:2002.08173}}.
\bibitem[{von Manteuffel and Schabinger(2015)}]{vonManteuffel:2014ixa}
\bibinfo{author}{A.~von Manteuffel}, \bibinfo{author}{R.~M. Schabinger},
\newblock \bibinfo{title}{{A novel approach to integration by parts
  reduction}},
\newblock \bibinfo{journal}{Phys. Lett. B} \bibinfo{volume}{744}
  (\bibinfo{year}{2015}) \bibinfo{pages}{101--104}.
  \DOIprefix\doi{10.1016/j.physletb.2015.03.029}.
  \href{http://arxiv.org/abs/1406.4513}{{\tt arXiv:1406.4513}}.
\bibitem[{Peraro(2016)}]{Peraro:2016wsq}
\bibinfo{author}{T.~Peraro},
\newblock \bibinfo{title}{{Scattering amplitudes over finite fields and
  multivariate functional reconstruction}},
\newblock \bibinfo{journal}{JHEP} \bibinfo{volume}{12} (\bibinfo{year}{2016})
  \bibinfo{pages}{030}. \DOIprefix\doi{10.1007/JHEP12(2016)030}.
  \href{http://arxiv.org/abs/1608.01902}{{\tt arXiv:1608.01902}}.
\bibitem[{Klappert and Lange(2020)}]{Klappert:2019emp}
\bibinfo{author}{J.~Klappert}, \bibinfo{author}{F.~Lange},
\newblock \bibinfo{title}{{Reconstructing rational functions with FireFly}},
\newblock \bibinfo{journal}{Comput. Phys. Commun.} \bibinfo{volume}{247}
  (\bibinfo{year}{2020}) \bibinfo{pages}{106951}.
  \DOIprefix\doi{10.1016/j.cpc.2019.106951}.
  \href{http://arxiv.org/abs/1904.00009}{{\tt arXiv:1904.00009}}.
\bibitem[{Peraro(2019)}]{Peraro:2019svx}
\bibinfo{author}{T.~Peraro},
\newblock \bibinfo{title}{{FiniteFlow: multivariate functional reconstruction
  using finite fields and dataflow graphs}},
\newblock \bibinfo{journal}{JHEP} \bibinfo{volume}{07} (\bibinfo{year}{2019})
  \bibinfo{pages}{031}. \DOIprefix\doi{10.1007/JHEP07(2019)031}.
  \href{http://arxiv.org/abs/1905.08019}{{\tt arXiv:1905.08019}}.
\bibitem[{Laurentis and Ma\^\i{}tre(2019)}]{Laurentis:2019bjh}
\bibinfo{author}{G.~Laurentis}, \bibinfo{author}{D.~Ma\^\i{}tre},
\newblock \bibinfo{title}{{Extracting analytical one-loop amplitudes from
  numerical evaluations}},
\newblock \bibinfo{journal}{JHEP} \bibinfo{volume}{07} (\bibinfo{year}{2019})
  \bibinfo{pages}{123}. \DOIprefix\doi{10.1007/JHEP07(2019)123}.
  \href{http://arxiv.org/abs/1904.04067}{{\tt arXiv:1904.04067}}.
\bibitem[{De~Laurentis and Page(2022)}]{DeLaurentis:2022otd}
\bibinfo{author}{G.~De~Laurentis}, \bibinfo{author}{B.~Page},
\newblock \bibinfo{title}{{Ans\"atze for scattering amplitudes from p-adic
  numbers and algebraic geometry}},
\newblock \bibinfo{journal}{JHEP} \bibinfo{volume}{12} (\bibinfo{year}{2022})
  \bibinfo{pages}{140}. \DOIprefix\doi{10.1007/JHEP12(2022)140}.
  \href{http://arxiv.org/abs/2203.04269}{{\tt arXiv:2203.04269}}.
\bibitem[{Magerya(2022)}]{Magerya:2022hvj}
\bibinfo{author}{V.~Magerya},
\newblock \bibinfo{title}{{Rational Tracer: a Tool for Faster Rational Function
  Reconstruction}}  (\bibinfo{year}{2022}).
  \href{http://arxiv.org/abs/2211.03572}{{\tt arXiv:2211.03572}}.
\bibitem[{Abreu et~al.(2019)Abreu, Dormans, Febres~Cordero, Ita, Page, and
  Sotnikov}]{Abreu:2019odu}
\bibinfo{author}{S.~Abreu}, \bibinfo{author}{J.~Dormans},
  \bibinfo{author}{F.~Febres~Cordero}, \bibinfo{author}{H.~Ita},
  \bibinfo{author}{B.~Page}, \bibinfo{author}{V.~Sotnikov},
\newblock \bibinfo{title}{{Analytic Form of the Planar Two-Loop Five-Parton
  Scattering Amplitudes in QCD}},
\newblock \bibinfo{journal}{JHEP} \bibinfo{volume}{05} (\bibinfo{year}{2019})
  \bibinfo{pages}{084}. \DOIprefix\doi{10.1007/JHEP05(2019)084}.
  \href{http://arxiv.org/abs/1904.00945}{{\tt arXiv:1904.00945}}.
\bibitem[{Heller and von Manteuffel(2022)}]{Heller:2021qkz}
\bibinfo{author}{M.~Heller}, \bibinfo{author}{A.~von Manteuffel},
\newblock \bibinfo{title}{{MultivariateApart: Generalized partial fractions}},
\newblock \bibinfo{journal}{Comput. Phys. Commun.} \bibinfo{volume}{271}
  (\bibinfo{year}{2022}) \bibinfo{pages}{108174}.
  \DOIprefix\doi{10.1016/j.cpc.2021.108174}.
  \href{http://arxiv.org/abs/2101.08283}{{\tt arXiv:2101.08283}}.
\bibitem[{Fieker et~al.(2017)Fieker, Hart, Hofmann, and Johansson}]{Nemo}
\bibinfo{author}{C.~Fieker}, \bibinfo{author}{W.~Hart},
  \bibinfo{author}{T.~Hofmann}, \bibinfo{author}{F.~Johansson},
\newblock \bibinfo{title}{Nemo/hecke: computer algebra and number theory
  packages for the julia programming language},
\newblock in: \bibinfo{booktitle}{Proceedings of the 2017 acm on international
  symposium on symbolic and algebraic computation}, \bibinfo{year}{2017}, pp.
  \bibinfo{pages}{157--164}.
\bibitem[{Johansson(2023)}]{Calcium}
\bibinfo{author}{F.~Johansson}, \bibinfo{title}{{Calcium}},
  \bibinfo{howpublished}{\url{https://github.com/fredrik-johansson/calcium}},
  \bibinfo{year}{2023}. \bibinfo{note}{Accessed: 2023-10-30}.
\bibitem[{Ruijl et~al.(2017)Ruijl, Ueda, and Vermaseren}]{Ruijl:2017dtg}
\bibinfo{author}{B.~Ruijl}, \bibinfo{author}{T.~Ueda},
  \bibinfo{author}{J.~Vermaseren},
\newblock \bibinfo{title}{{FORM version 4.2}}  (\bibinfo{year}{2017}).
  \href{http://arxiv.org/abs/1707.06453}{{\tt arXiv:1707.06453}}.
\bibitem[{Bern et~al.(2023)Bern, Herrmann, Roiban, Ruf, Smirnov, Smirnov, and
  Zeng}]{Bern:2023ccb}
\bibinfo{author}{Z.~Bern}, \bibinfo{author}{E.~Herrmann},
  \bibinfo{author}{R.~Roiban}, \bibinfo{author}{M.~S. Ruf},
  \bibinfo{author}{A.~V. Smirnov}, \bibinfo{author}{V.~A. Smirnov},
  \bibinfo{author}{M.~Zeng},
\newblock \bibinfo{title}{{Conservative binary dynamics at order $O(\alpha^5)$
  in electrodynamics}}  (\bibinfo{year}{2023}).
  \href{http://arxiv.org/abs/2305.08981}{{\tt arXiv:2305.08981}}.

\end{thebibliography}
\end{document}